\documentclass[a4paper]{jpconf}
\usepackage{graphicx,amssymb,amsfonts,amsmath}

\def\Eo{\,{\mathbb E}}
\def\Ro{\,{\mathbb R}}
\newcommand{\be}{\begin{eqnarray}}
\newcommand{\ee}{\end{eqnarray}}

\def\qq{{\bf q}}
\def\pp{{\bf p}}

\begin{document}
\title{The $q$-exponential family in statistical physics}

\author{Jan Naudts}

\address{Universiteit Antwerpen, Groenenborgerlaan 1717, 2020 Antwerpen, Belgium}

\ead{jan.naudts@ua.ac.be}

\begin{abstract}
The Boltzmann-Gibbs probability distribution, seen as a statistical model, belongs to the
exponential family. Recently, the latter concept has been generalized. The $q$-exponential family
has been shown to be relevant for the statistical description of small isolated systems.
Two main applications are reviewed: 1. The distribution of the momentum of a single particle
is a $q$-Gaussian, the distribution of its velocity is a deformed Maxwellian; 2.
The configurational density distribution belongs to the $q$-exponential family.

The definition of the temperature of small isolated systems is discussed. It depends
on defining the thermodynamic entropy of a microcanonical ensemble in a suitable manner.
The simple example of non-interacting harmonic oscillators shows that R\'enyi's entropy
functional leads to acceptable results.
\end{abstract}

\section{Introduction}

In statistics, a model is a probability distribution which depends on a number of parameters.
The probability distributions of statistical physics usually depend
on parameters and can therefore be seen as statistical models.
Typical parameters are the total energy $U$ in the microcanonical ensemble, or the inverse temperature $\beta$
in the canonical ensemble. This parameter dependence is important
in order to understand why certain models belong to the exponential family
and others do not. In particular, all models described by a Boltzmann-Gibbs distribution belong to the
exponential family because they have the right dependence on the inverse temperature $\beta$.

Recently, the notion of the exponential family has been generalized by the present author in a series of papers
\cite {NJ04d,NJ05b,NJ05,NJ06,NJ08,NJ08b}.
The same definition of generalized exponential family has been introduced in the mathematics literature
\cite {NJ04d,GD04,ES04,SV07,NJ08}.
This class of models was also derived using the maximum entropy principle in \cite {AS03b,HT07}
and in the context of game theory \cite {TF07,TF09}.
Here we concentrate on the more specific notion of the $q$-exponential family and review the
evidence that this family appears in a natural manner in the context of the microcanonical description
of a classical gas.

The notion of $q$-exponential family is connected with Amari's $\alpha$-family \cite {AS85}, studied
in the context of information geometry. The geometric approach is very appealing also in the context
of statistical physics. See for instance \cite {OA07,OA09}. However, this topic will not be discussed in the
present paper.

Note that in many of the formulas given below Boltzmann's constant
$k_{\rm B}$ is set equal to 1, which means that temperatures are measured in units of energy.

\section{The Boltzmann-Gibbs distribution}

The Boltzmann-Gibbs (BG) distribution is given by
\be
 f_\beta(x)=\frac {c(x)}{Z(\beta)}e^{-\beta H(x)},
\qquad x \mbox { in phase space}
\label {BG:bg}
\ee
The {\sl Hamiltonian} $H(x)$ is a given function. It determines the {\sl physical model}.
$\beta$ is a parameter, corresponding with the inverse temperature of thermodynamics.
The function $c(x)$ quite often is just a constant.
$Z(\beta)$ is called the {\sl partition sum}.
It normalizes the probability distribution:
\be
Z(\beta)=\int{\rm d}x\,c(x) e^{-\beta H(x)}.
\ee
$\Phi(\beta)\equiv\ln Z(\beta)$ 
corresponds in thermodynamics with Massieu's function.
It satisfies the identity
\be
\frac{{\rm d} \Phi}{{\rm d}\beta}=-U
\label {BG:Phider}
\ee
with $U$ the {\sl energy}, this is, the expected value of the Hamiltonian $H(x)$
\be
U=\Eo_\beta H=\int{\rm d}x\,f_\beta(x)H(x).
\ee

The second derivative of Massieu's function satisfies
\be
\frac{{\rm d}^2 \Phi}{{\rm d}\beta^2}=\Eo_\beta(H-U)^2\ge 0.
\ee
Therefore, $\Phi(\beta)$ is a convex function, as expected from thermodynamics.
Indeed, in thermo\-dynamics Massieu's function is by definition \cite {CHB85}
the Legendre transform of the {\sl thermodynamic entropy}
$S(U)$, which is a function of the energy $U$. Within statistical physics
the thermodynamic entropy can be defined by means of an {\sl entropy functional},
$I(f)$, which is a function over the space of all probability distributions $f$.
The most common entropy functional is that of Boltzmann-Gibbs-Shannon (BGS)
\be
 I(f)=-\int{\rm d}x\,f(x)\ln\frac {f(x)}{c(x)}.
\label {BG:entropy}
\ee
It is well-known that $I(f)$ takes its maximal value at $f=f_\beta$ within
the set of all $f$ satisfying $\Eo_fH=U$. This maximal value of $I(f)$
is then identified with the thermodynamic entropy $S(U)$. It satisfies
\begin{eqnarray}
 S(U)&\equiv&\max_f\{I(f):\,\Eo_fH=U\}=I(f_\beta)\crcr
&=&\int{\rm d}x\,f_\beta(x)
\left(\ln Z(\beta)+\beta H(x)
\right)\crcr
&=&\ln Z(\beta)+\beta U.
\label {BG:thermentr}
\end{eqnarray}
Comparison with
\be
 \Phi(\beta)=\max_U\{S(U)-\beta U\}
\label {BG:phisrel}
\ee
then yields $\Phi(\beta)=\ln Z(\beta)$.
From (\ref {BG:thermentr}, \ref {BG:phisrel}) follows the inverse
transformation
\be
 S(U)=\inf_\beta\{\Phi(\beta)+\beta U\}
\label {BG:inftrans}
\ee
and an identity, which is the dual expression of (\ref {BG:Phider}),
\be
\beta=\frac {{\rm d}S}{{\rm d}U}.
\label {BG:invtemp}
\ee
The latter relation is often used as the definition of the inverse temperature.
Another consequence of (\ref {BG:inftrans}) is that $S(U)$ is
a concave function.

\section{The exponential family}

A model of classical (statistical) mechanics is determined by a Hamiltonian $H(x)$.
On the other hand, a 
{\sl statistical model} is defined as a parameterized probability distribution function (pdf).
It is now immediately clear that the Boltzmann-Gibbs distribution (\ref {BG:bg})
defines a statistical model with one parameter $\beta$. Moreover, it belongs
to a very special class of models called {\sl the exponential family}.
In fact, a model belongs to the exponential family with parameters
$\theta=(\theta_1,\theta_2,\cdots,\theta_n)$ when
its pdf $f_\theta(x)$ can be written into the canonical form
\be
 f_\theta(x)=c(x)\exp\left(\sum_{j=1}^n\theta_jK_j(x)-\alpha(\theta)\right),
\label {expfam:def}
\ee
where $c(x)$, $K_j(\theta)$, and $\alpha(\theta)$ are known functions.

Models belonging to the exponential family share a number of nice properties.
For instance, they satisfy the identities, which generalize (\ref {BG:Phider}),
\be
\frac {\partial\alpha}{\partial\theta_j}=\Eo_\theta K_j=\int{\rm d}x\,f_\theta(x) K_j(x).
\ee
Note that the functions $K_j(x)$ may not depend on the parameters $\theta$.
In statistical physics one uses sometimes effective Hamiltonians which
depend on the temperature $\beta^{-1}$. The corresponding statistical models may exhibit thermodynamic instabilities which cannot occur in models belonging to the exponential family.

\section{Deformed logarithmic and exponential functions}

The Boltzmann-Gibbs distribution is so dominant in statistical physics
that one can get the impression that this is the only relevant statistical model.
However there exists a much larger class of models, sharing most of the nice
properties of the exponential family. Some of these more general models appear in a natural manner
in statistical physics, as will be shown further on.
One way to get access to this larger class is by replacing the exponential
function in the definition (\ref {expfam:def}) by some other function.
This can be done in a fairly general way \cite {NJ04d,HT07}. Here, a one-parameter
deformation is considered, which is the basis for the generalization
known as {\sl non-extensive statistical mechanics} \cite {TC09}.

Fix a real number $q\not=1$. The $q$-deformed logarithm is defined by \cite {TC94,NJ02}
\be
\ln_q(u)=\frac 1{1-q}\left(u^{1-q}-1\right),
\qquad u>0.
\ee
The basic properties of the $q$-deformed logarithmic function are that $\ln_q(1)=0$ and 
that it is  a strictly increasing function. Indeed,
its first derivative is always positive
\be
\frac {{\rm d}\,}{{\rm d}u}\ln_q(u)=\frac 1{u^q}>0
\ee
For $q>0$ the $q$-logarithm is a concave function.
In the limit $q=1$ it reduces to the natural logarithm $\ln u$.

\begin{figure}[h]
\begin{minipage}{14pc}
\includegraphics[width=14pc]{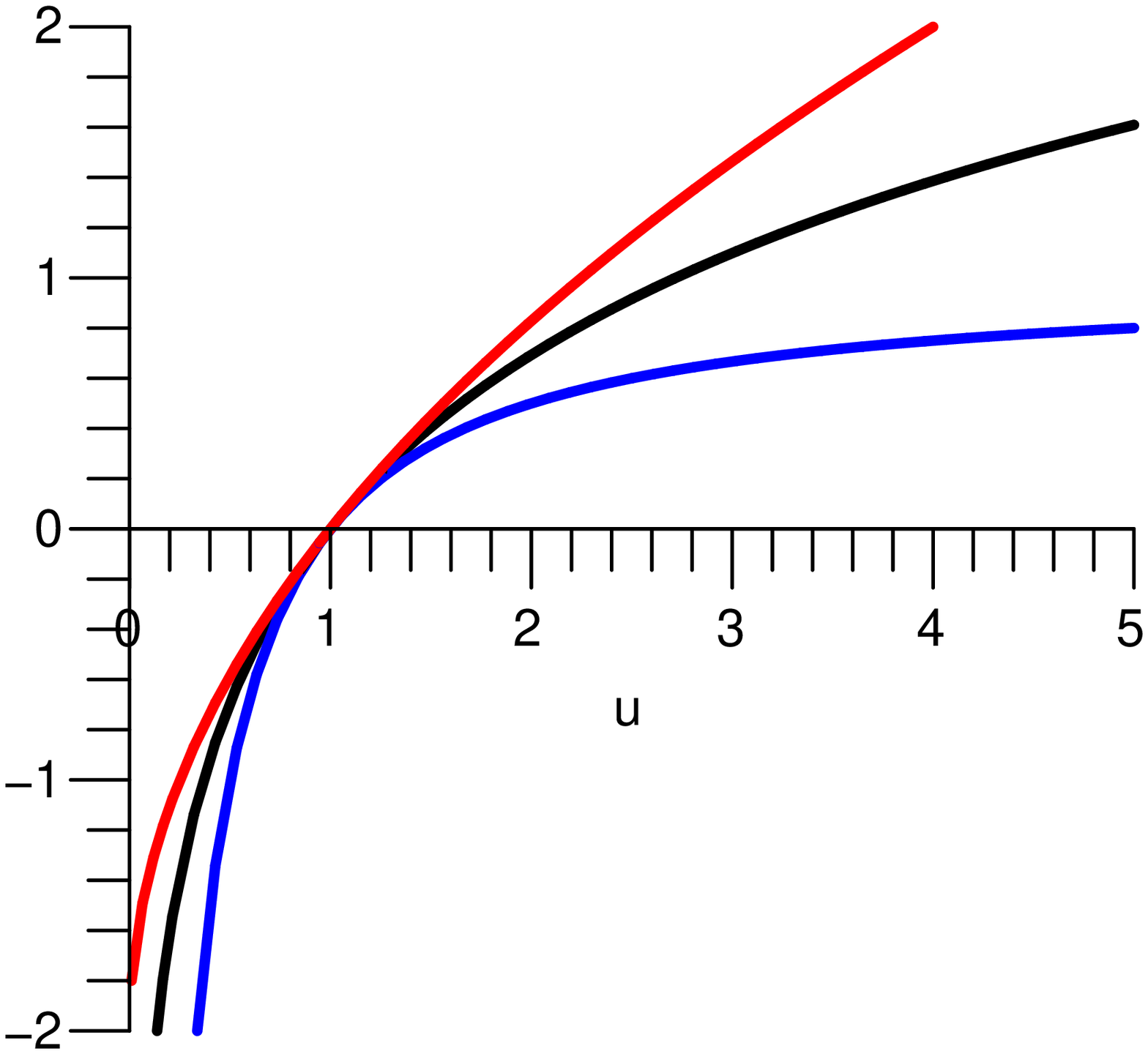}
\caption{\label{fig:qlog}
The $q$-deformed logarithm $\ln_q(u)$ for $q$-values $0.5$ (red), $1$ (black), and $2$ (blue).}
\end{minipage}\hspace{2pc}%
\begin{minipage}{14pc}
\includegraphics[width=14pc]{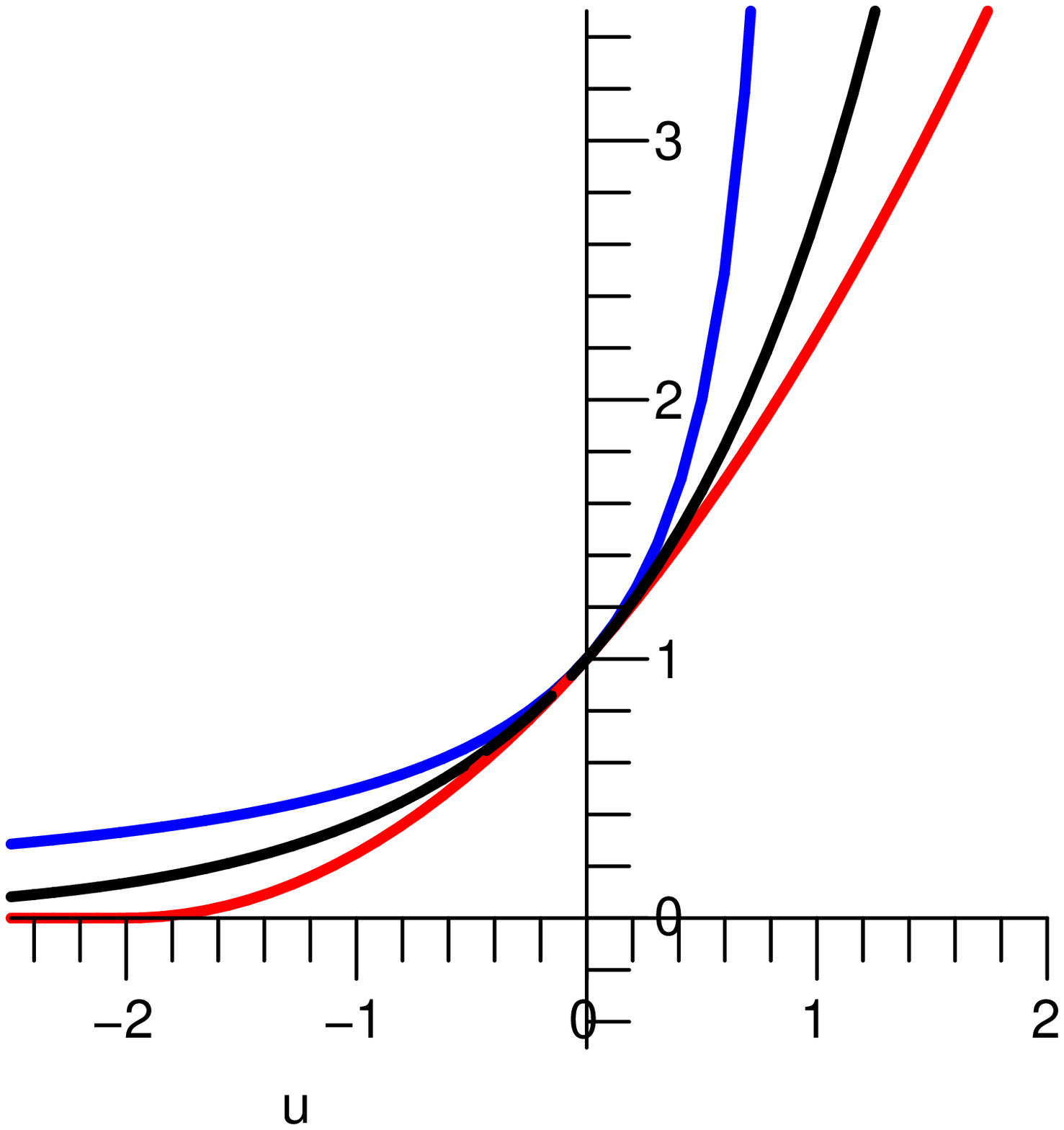}
\caption{\label{fig:qexp}
The $q$-deformed exponential $\exp_q(u)$ for $q$-values $0.5$ (red),$1$ (black), and $2$ (blue).}
\end{minipage} 
\end{figure}

The inverse function is the deformed exponential function
\be
\exp_q(u)=\left[1+(1-q)u\right]_+^{1/(1-q)}.
\ee
The notation $[u]_+=\max\{0,u\}$ is used. One has $0\le\exp_q(u)\le+\infty$ for all $u$.
For $q\not=1$ the range of $\ln_q(u)$ is not the full line. By putting $\exp_q(u)=0$ when
$u$ is below the range of $\ln_q(u)$, and equal to $+\infty$ when it is above,
$\exp_q(u)$ is an increasing function of $u$, defined for all values of $u$.
With this extended definition one has
\be
\exp_q(\ln_q(u))=u
\quad\mbox{ for all }u>0.
\ee
However, the relation $\ln_q(\exp_q(u))=u$ is only defined when $\exp_q(u)$ is neither 0 or $+\infty$.

The basic properties of the $q$-deformed exponential are that $\exp_q(0)=1$
and that it is an increasing function. Indeed, one has
\be
\frac {{\rm d}\,}{{\rm d}u}\exp_q(u)=\left[\exp_q(u)\right]^q\ge 0.
\ee
For $q>0$ it is a convex function.

\section{The $q$-exponential family}

The deformed logarithmic and exponential functions can be used in more than one way
to develop a generalized formalism of statistical physics.
The original approach of non-extensive statistical physics \cite {TC88}
was to first generalize the Boltzmann-Gibbs-Shannon entropy
functional (\ref {BG:entropy}) by deforming the logarithm contained in it.
Next one studies pdfs which maximize this entropy functional under given constraints.
However, it turns out to be advantageous to generalize immediately
the notion of exponential family by deforming the exponential function appearing in (\ref {expfam:def}).
This leads to the definition that the one-parameter model $f_\theta(x)$ belongs
to the {\sl $q$-exponential family} if there exist functions $c(x),\alpha(\theta),H(x)$
such that one can write
\be
f_\theta(x)=c(x)\exp_q\left(-\alpha(\theta)-\theta H(x)\right).
\label {qexp:qexpfam}
\ee
In the limit $q=1$ this reduces to the standard definition of the exponential family
with $\theta=\beta$. Note that we take into account the possibility that the parameter
$\theta$ does not necessarily coincide with the inverse temperature $\beta$
as it is defined by the thermodynamic relation (\ref {BG:invtemp}).

Introduce the $q$-deformed entropy functional
\be
I_q(f)&=&-\int{\rm d}x\,f(x)\ln_q\frac {f(x)}{c(x)}\crcr
&=&\frac 1{1-q}\left(1-\int{\rm d}x\,c(x)\left(\frac {f(x)}{c(x)}\right)^{2-q}
\right).
\label {qexp:entropy}
\ee
In the limit $q=1$ it reduces to the BGS-entropy functional (\ref {BG:entropy}).
Note that this definition is almost the entropy functional
originally introduced by Tsallis \cite {TC88}, but differs from it by the
fact that the parameter $q$ has been substituted by $2-q$.

Assume now $0<q<2$. Then one can show that the pdf  $f=f_\theta(x)$ maximizes the quantity
\be
\frac 1{2-q}I_q(f)-\theta\Eo_f H.
\label {qexp:varprin}
\ee
In the $q=1$-case this result is known as {\sl the variational principle}.
A proof of this statement involves the introduction of an appropriate Bregman divergence
--- see \cite {NJ04d,NJ08}.
Note that the variational principle implies {\sl the maximum entropy principle}:
$I_q(f_\theta)\ge I_q(f)$ for all $f$ which satisfy $\Eo_f H=\Eo_\theta H$
(using the notation $\Eo_\theta\equiv\Eo_{f_\theta}$).
A well-known problem is now that $f_\theta$ also maximizes $\xi(I_q(f))$,
where $\xi(u)$ is an arbitrary monotonically increasing function.
Therefore a meaningful {\sl ansatz} is to assume that
the thermodynamic entropy $S(U)$ is given by
\be
S(U)=\xi\left(I_q(f_\theta)\right)=\xi\left(\alpha(\theta)+\theta\Eo_\theta H\right).
\ee
The thermodynamic expression for the inverse temperature $\beta$
can then be calculated using (\ref {BG:invtemp}).
It is clear that the resulting relation between energy $U$ and temperature $\beta^{-1}$
will depend on the choice of the monotonic function $\xi(u)$. Physical arguments are needed
to decide which choice is the right one.

\section{$q$-Gaussians}

The $q$-Gaussian distribution in one variable is given by 
(see for instance \cite {VP07,VP07b,OW08})
\be
f(x)=\frac {1}{c_q \sigma}\exp_q(- x^2/\sigma^2),
\label {eg:qgauss}
\ee
with
\be
c_q=\int_{-\infty}^\infty{\rm d}x\,\exp_q(- x^2)
&=& \sqrt{\frac {\pi}{q-1}}\frac{\Gamma\left(-\frac 12+\frac 1{q-1}\right)}{\Gamma\left(\frac 1{q-1}\right)}
\quad \mbox { if }\quad 1<q<3,\crcr
&=& \sqrt{\frac {\pi}{1-q}}\frac{\Gamma\left(1+\frac 1{1-q}\right)}{\Gamma\left(\frac 32+\frac 1{1-q}\right)}
\quad \mbox { if } \quad q<1.
\ee
It can be brought into the form (\ref {qexp:qexpfam}) with
$c(x)=1/c_q$, $H(x)=x^2$, $\theta=\sigma^{3-q}$, and
\be
\alpha(\theta)=\frac {\sigma^{q-1}-1}{q-1}=\ln_{2-q}(\sigma).
\ee

The $q=1$-case reproduces the conventional Gauss distribution.
For $q<1$ the distribution vanishes outside an interval.
Take for instance $q=1/2$. Then (\ref{eg:qgauss}) becomes
\be
f(x)&=&\frac {15\sqrt 2}{32\sigma}\left[1-\frac {x^2}{\sigma^2}\right]_+^2.
\label {qgauss:onehalf}
\ee
This distribution vanishes outside the interval $[-\sigma,\sigma]$.
In the range $1\le q< 3$ the $q$-Gaussian is strictly positive on the whole
line and decays with a power law in $|x|$ instead of exponentially.
For $q=2$ one obtains
\be
f(x)=\frac 1\pi \frac {\sigma}{x^2+\sigma^2}.
\label {cauchy}
\ee
This is the Cauchy distribution. The function (\ref {cauchy})
is also called a Lorentzian
and is often used in physics to fit the shape of spectral lines.
For $q\ge 3$ the distribution cannot be normalized because
\be
f(x)\sim\frac 1{|x|^{2/(q-1)}}
\mbox{ as }|x|\rightarrow\infty
\ee
is not integrable anymore.

\begin{figure}[h]
\includegraphics[width=14pc]{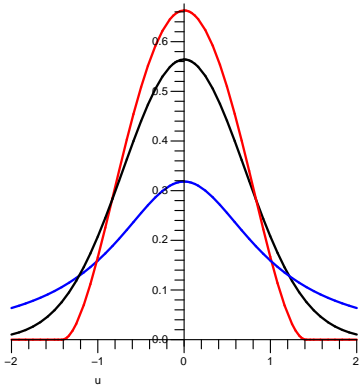}\hspace{2pc}%
\begin{minipage}[b]{14pc}\caption{\label{fig:qgauss}
$q$-Gaussians for $\sigma=1$ and for $q$-values $0.5$ (red), $1$ (black), and $2$ (blue).}
\end{minipage}
\end{figure}

\section{A $q$-deformed Maxwell distribution}

The most straightforward way to produce an important example of $q$-Gaussians in statistical physics is
by considering marginals of the uniform distribution on a hypersphere -- see \cite {VP05}.
The Hamiltonian of an $N$-particle classical ideal gas is given by
\be
H(p)=\frac 1{2m}\sum_{j=1}^N|\pp_{j}|^2,
\ee
where $m$ is the mass of the particles and ${\pp}_j$ is the momentum
of the $j$-th particle. Given the value $E$ for the total energy, the phase space
consists of all points of the $3N$-dimensional sphere with radius $\sqrt {2mE}$.
Assume equal probability of all points of phase space.
Let $B_n(r)$ denote the volume of a sphere with radius $r$ in dimension $n$.
The probability distribution for the momentum of a single particle becomes
\be
f({\pp}_{1})
&=&\frac 1{(2m)^{3N/2}B'_{3N}(E)}\int{\rm d}^3{\pp}_2\cdots{\rm d}^3{\pp}_N\,\delta(E-H(p))\crcr
&=&\frac 1{(2m)^{3/2}}\,\frac {B'_{3(N-1)}(E-|{\pp}_1|^2/2m)}{B'_{3N}(E)}\crcr
&=&\frac 1{c_q\sigma}\exp_q(-|{\pp}_1|^2/2\sigma^2)
\label {maxw:pdistr}
\ee
with
\be
q=\frac {3N-6}{3N-4}
\quad\mbox{ and }\quad
\sigma^2=\frac {2mE}{3N-4}.
\ee
This is a $q$-Gaussian with $q<1$. The appearance of a cutoff can be easily understood.
Arbitrary large momenta are not possible because of the obvious upperbound $|{\pp}_1|^2\le 2mE$.
The probability distribution of the scalar velocity $v=|{\pp}_1|/m$ becomes
\be
f_q(v)=\frac {4\pi m^3}{c_q\sigma}v^2\exp_q\left(-\frac 1{2\sigma^2}mv^2\right).
\ee
Only in the limit of large systems it converges to the Maxwell distribution.
One concludes that the Maxwell distribution is an approximation and is only
valid in the limit $N\rightarrow\infty$.

\section{Configurational density of classical gases}

The configurational density of a gas of classical particles
always belongs to the $q$-exponential family \cite {NB09}.
The idea behind the calculation is the same as that of the
previous section. This kind of non-extensivity of a classical gas
was observed already quite some time ago \cite {PP94,CM07}.
It becomes a strong statement in the language of $q$-exponential families.

A classical gas of $N$ interacting particles is described by the Hamiltonian
\be
H(\qq,\pp)=
\frac 1{2m}\sum_{j=1}^N|\pp_j|^2+{\cal V}(\qq),
\ee
where ${\cal V}(\qq)$ is the potential energy due to interaction
among the particles and between the particles and the walls of the system.
The microcanonical ensemble is then by definition the following singular probability distribution
\be
f_U(\qq,\pp)=\frac 1{\omega(U)}\delta(U-H(\qq,\pp)),
\ee
where $\delta(\cdot)$ is Dirac's delta function and $\omega(U)$ is the normalization
\be
\omega(U)=\frac 1{h^{3N}}\int_{\Ro^{3N}}{\rm d}\pp_1\cdots{\rm d}\pp_N
\int_{\Ro^{3N}}{\rm d}\qq_1\cdots{\rm d}\qq_N\,\delta(U-H(\qq,\pp)).
\label {cpd:densstate}
\ee
The constant $h$ is introduced for dimensional reasons.
Because of the quadratic nature of the kinetic energy term it is possible to
integrate out the momenta ${\pp}_j$. The resulting function
is called the {\sl configurational density function} and is evaluated as follows
\be
f_U^{\rm conf}(\qq)
&=&\frac {1}{h^{3N}}\int_{\Ro^{3N}}{\rm d}\pp_1\cdots{\rm d}\pp_N\,f_U(\qq,\pp)\crcr
&=&\frac {1}{h^{3N}}\frac 1{\omega(U)}\int_{\Ro^{3N}}{\rm d}\pp_1\cdots{\rm d}\pp_N\,\delta(U-H(\qq,\pp))\crcr
&=&\frac {1}{h^{3N}}\frac 1{\omega(U)}\frac {{\rm d}\,}{{\rm d}U}
\int_{\Ro^{3N}}{\rm d}\pp_1\cdots{\rm d}\pp_N\,\Theta\left(U-{\cal V}(\qq)-\frac 1{2m}\sum_{j=1}^N|\pp_j|^2\right)\crcr
&=&\frac {1}{h^{3N}}\frac 1{\omega(U)}(2m)^{3N/2}B(3N)
\frac {{\rm d}\,}{{\rm d}U}
[U-{\cal V}(\qq)]_+^{3N/2}\crcr
&=&\frac {1}{2h^{3N}}\frac {3N}{\omega(U)}(2m)^{3N/2}B(3N)
\left[U-{\cal V}(\qq)\right]_+^{\frac 32N-1}\crcr
&=&c_N\exp_q\left(-\alpha(\theta)-\theta {\cal V}(\qq)\right),
\label {cpd:pdf}
\ee
with
\be
c_N&=&\left(\frac {2m}{h^2}\right)^{3N/2},\qquad
\theta=\frac 1{1-q}\frac 1{[\Gamma(3N/2)\omega(U)]^{1-q}},\crcr
\alpha(\theta)&=&\frac 32N-1-\theta U,\qquad
q=1-\frac 2{3N-2}.
\label {cpd:params}
\ee
Assume now that $\omega(U)$ is a strictly increasing function. Then
it can be inverted to obtain $U$ as a function of $\theta$.
Under this assumption the configurational density function belongs
to the $q$-exponential family.
As a consequence, one immediately knows that $f_U^{\rm conf}(\qq)$
maximizes the entropy functional $I_q(f)$. It is then tempting
to identify $I_q(f_U^{\rm conf})$ with the configurational entropy.
However, as noted before, $\xi(I_q(f_U^{\rm conf}))$ could serve as well,
where $\xi(u)$ is an arbitrary monotonic function.

\section{On the definition of temperature}

There is no consensus in the literature what is the correct definition of
the thermodynamic entropy $S(U)$ for isolated systems. The question
is of importance because it directly determines the definition of
the thermodynamic temperature via the relation (\ref {BG:invtemp}).
Most often one quotes the definition known as Boltzmann's entropy
\be
S(U)=k_B\ln \omega(U),
\ee
where $\omega(U)$ is the density of states (\ref {cpd:densstate}).
However, this choice of definition of entropy has some drawbacks.
For instance, for the pendulum the entropy $S(U)$ as a function of
internal energy $U$ is a piecewise convex function
instead of a concave function \cite {NJ05b}.
The lack of concavity can be interpreted as a microcanonical instability \cite {GDH01,GD90}.
But there is no physical reason why the pendulum should be classified as being instable
at all energies.

The shortcomings of Boltzmann's entropy have been noticed long ago.
A slightly different definition of entropy is \cite {CM07,SA48,PHT85,CB07}
(see also in \cite {SBJ06} the reference to the work of A. Schl\"uter )
\be
S(U)=k_B\ln\Omega(U),
\label {temp:pearson}
\ee
where $\Omega(U)$ is the integral of $\omega(U)$ and is given by
\be
\Omega(U)=\frac 1{h^{3N}}\int_{\Ro^{3N}}{\rm d}\pp_1\cdots{\rm d}\pp_N
\int_{\Ro^{3N}}{\rm d}\qq_1\cdots{\rm d}\qq_N
\Theta(U-H(\qq,\pp)).
\ee
An immediate advantage of (\ref {temp:pearson}) is that
the resulting expression for the temperature $T$, 
defined by the thermodynamical formula
\be
\frac 1{T}=\frac {{\rm d}S}{{\rm d}U},
\label {temp:Tdef}
\ee
coincides with the notion of temperature as used by experimentalists.
Indeed, one finds
\be
k_BT=\frac {\Omega(U)}{\omega(U)}.
\label {temp:Tres}
\ee
For a harmonic oscillator the density of states $\omega(U)$ is a constant.
Hence, (\ref {temp:Tres}) implies $k_BT=U$, as wanted.
It is well-known
that for classical monoatomic gases the r.h.s.~of (\ref {temp:Tres}) coincides
with twice the average kinetic energy per degree of freedom.
Its significance is that the
equipartition theorem, assigning $(1/2)k_BT$ to each degree of
freedom, does hold for the kinetic energy also in the microcanonical ensemble.
Quite often the average kinetic energy per degree of freedom
is experimentally accessible and provides a unique way to measure
accurately the temperature of the system.

But also (\ref {temp:pearson}) and (\ref {temp:Tres}) are subject to criticism.
In small systems finite size corrections appear \cite {SBJ06,USC08} for a number
of reasons. As argued in \cite {USC08}, the problem is not the equipartition of the
kinetic energy over the various degrees of freedom, but the relation between
temperature and kinetic energy.

\section{Example}

Let us follow \cite {NB09} and consider the example of
$3N$ non-interacting harmonic oscillators.
The potential energy equals
\be
{\cal V}(\qq)=\frac12m \omega^2\sum_{j=1}^{3N}\qq_j^2.
\ee
One calculates
\be
\Omega(U)&=&\frac 1{h^{3N}}\int_{\Ro^{3N}}{\rm d}\pp_1\cdots{\rm d}\pp_{3N}
\int_{\Ro^{3N}}{\rm d}\qq_1\cdots{\rm d}\qq_{3N}
\Theta\left(U-\frac 1{2m}\sum_j\pp_j^2-\frac 12m\omega^2\sum_j\qq_j^2
\right)\crcr
&=&\frac 1{(3N)!\,\omega^{3N}}\left(\frac {4\pi U}{h}\right)^{3N}.
\label {ex:Omega}
\ee
From (\ref {cpd:params}) and (\ref {ex:Omega}) now follows
\be
\frac 1\theta=(1-q)\left[
\frac {\Gamma(3N/2)}{\Gamma(3N)}\frac 1{\omega^{3N}}\left(\frac {4\pi }{h}\right)^{3N}
U^{3N-1}
\right]^{1-q}.
\ee
Using Stirling's approximation one obtains
\be
\frac {\Gamma(3N/2)}{\Gamma(3N)}
\sim\sqrt{2}\left(\frac {\rm e}{6N}\right)^{3N/2}
\ee
so that
\be
\frac 1\theta\sim {\rm e}\left(\frac {4\pi}{h\omega}\frac U{3N}\right)^2,
\label {ho:restemp}
\ee
which shows that $1/\theta$ is roughly proportional to $U^2$.
This is not what one expects. In the canonical ensemble is
the temperature $1/\beta$  proportional to $U$
\be
\frac 1\beta=\frac U{3N}.
\label {ho:temp}
\ee
One therefore concludes that the parameter $\theta$ cannot be
the inverse temperature $\beta$. Note that
$\theta={\rm d}\tilde S^{\rm conf}/{{\rm d}U^{\rm conf}}$
holds with $\tilde S^{\rm conf}\equiv I_q(f^{\rm conf}_U)$
 and $U^{\rm conf}=\Eo_\theta{\cal V}=\frac 12 U$. Therefore
one concludes that
$\tilde S^{\rm conf}$ cannot be the thermodynamic configurational entropy $S^{\rm conf}$,
but the two entropies must be related by a non-trivial function $\xi(u)$.
From (\ref {ho:temp}) follows
\be
S^{\rm conf}=\frac {3N}2\ln U^{\rm conf}+A,
\ee
for some constant $A$. But (\ref {ho:restemp}) implies that
\be
\tilde S^{\rm conf}\sim -\frac 1{{\rm e}}\left(\frac {h\omega}{4\pi}\right)^2\frac {(3N)^2}{U^{\rm conf}}+
\mbox{constant.}
\label {ex:sappr}
\ee
Therefore the relation between $S^{\rm conf}$ and $\tilde S^{\rm conf}$ is logarithmic.
More precisely, $S^{\rm conf}=\xi(\tilde S^{\rm conf})$, with $\xi(x)$ of the form
 $\xi(x)=-3N\ln(B-x)+C$, with constants $B$ and $C$.
This suggests that R\'enyi's entropy functional is the right one to start with.
Indeed, let $\alpha=2-q$. The relation between R\'enyi's $I_\alpha(f)$ and $I_q(f)$, as given by 
(\ref {qexp:entropy}), is $I_\alpha(f)=\xi(I_q(f))$ with
\be
\xi(u)=-\frac 1{1-q}\ln\left(1-(1-q)u\right).
\ee
Take the constant in (\ref {ex:sappr}) equal to $1/(1-q)$. Then one obtains
\be
S^{\rm conf}\equiv
\xi\left(\tilde S^{\rm conf}\right)\simeq\left(\frac {3N}2-1\right)
\left[\ln \frac {U^{\rm conf}}{6N}
+\mbox{ constant}\right].
\ee
This yields
\be
\frac {{\rm d}S^{\rm conf}}{{\rm d}U^{\rm conf}}
=\left(\frac {3N}2-1\right)\frac 1{U^{\rm conf}}=\frac {3N-2}U,
\ee
which is an acceptable expression for the inverse temperature $\beta$.

\section{Concluding remarks}

The notion of a $q$-exponential family of statistical models has been reviewed.
The $q$-Gaussian distribution belongs to this family.
The probability distribution of the momentum of a single particle
in a classical gas is such a $q$-Gaussian. An immediate consequence is that
the velocity distribution of a classical particle is a $q$-deformed Maxwell
distribution. In particular, the standard Maxwell distribution is an approximation which is only
valid in the limit of large systems.

The configurational probability distribution of a classical gas
always belongs to the $q$-exponential family. The non-extensivity
parameter $q$ is given by
\be
\frac 1{1-q}=\frac 32N-1,
\label {disc:nep}
\ee
where $N$ is the number of particles. The latter expression has
appeared quite often in the literature, see for instance \cite {PP94,AMPP01,AMAA03}.

Finally, a system of $3N$ independent harmonic oscillators is considered.
It is known that Tsallis' entropy and R\'enyi's entropy are equivalent in
the context of the maximal entropy principle. But they make a difference
from a thermodynamical point of view. The example suggests that
the correct thermodynamical definition of temperature requires the use
of R\'enyi's entropy. This observation has been made before in a different
setting \cite {CM07,CB07}.

\section*{References}

{
\raggedright
\bibliographystyle{iopart-num.bst}
\bibliography{biblionaudts}
}

\end{document}